\documentclass[prd, aps, superscriptaddress, preprintnumbers, twocolumn, floatfix, nofootinbib]{revtex4}

\usepackage{amsfonts}
\usepackage{amsmath}
\usepackage{amssymb}
\usepackage{bm}
\usepackage{dcolumn}
\usepackage{graphicx}
\usepackage[latin1]{inputenc}
\usepackage{latexsym}
\usepackage{rotating}
\usepackage{hyperref}
\usepackage{graphicx}
\usepackage{color}
\usepackage{ulem}

\usepackage{amsfonts}
\usepackage{amsmath}
\usepackage{amssymb}
\usepackage{bm}
\usepackage{dcolumn}
\usepackage{graphicx}
\usepackage[latin1]{inputenc}
\usepackage{latexsym}
\usepackage{rotating}
\usepackage{hyperref}

\def\beqra{\begin{eqnarray}}
\def\eeqra{\end{eqnarray}}
\def\beq{\begin{equation}}
\def\eeq{\end{equation}}

\def\bx{{\bf{x}}}
\def\bk{{\bf{k}}}
\def\bp{{\bf{p}}}
\def\bq{{\bf{q}}}

\def\bv{{\bf{v}}}
\def\bd{{\bf{d}}}

\def\bV0{{\bf{V_0}}}
\def\re#1{(\ref{#1})}

\def\alt{~\mbox{\raisebox{-.6ex}{$\stackrel{<}{\sim}$}}~}
\def\bx{{\bf{x}}}

\def\br{{\bf{r}}}

\def\bk{{\bf{k}}}
\def\bp{{\bf{p}}}
\def\bq{{\bf{q}}}

\def\bv{{\bf{v}}}

\def\bz{{\bf{z}}}

\begin{document}

\preprint{YITP-19-70}

\title{Measuring Bias via the Consistency Relations of the Large Scale Structure }

\author{Marco Marinucci$^{1,2}$, Takahiro Nishimichi$^3$, and Massimo Pietroni$^{1,4}$}
\affiliation{%
$^1$Dipartimento di Scienze Matematiche, Fisiche ed Informatiche dell'Universit\`a di Parma, Italy \\
$^2$ INFN, Gruppo collegato di Parma, Italy\\
$^3$ Center for Gravitational Physics, Yukawa Institute for Theoretical Physics, Kyoto University, Kyoto 606-8502, Japan\\
$^4$ INFN, Sezione di Padova, Italy
}%
\begin{abstract}
Consistency Relations (CR) for the Large Scale Structure are exact equalities between correlation functions of different order. These relations descend from the equivalence principle and hold for primordial perturbations generated by single-field models of inflation. They are not affected by nonlinearities and hold  also for biased tracers and in redshift space. 
We show that Baryonic Acoustic Oscillations (BAO) in the bispectrum (BS) in the squeezed limit are suppressed with respect to those in the power spectrum (PS) by a  coefficient that depends on the BS configuration and on the bias parameter (and, in redshift space, also on the growth rate). We test these relations using large volume N-body simulations and show that they provide a novel way to measure large scale halo bias and, potentially, the growth rate. Since bias is obtained by comparing two directly observable quantities, the method is free from theoretical uncertainties both on the computational scheme and on the underlying cosmological model. 
\end{abstract}


\maketitle


\section{\label{intro}Introduction}
The Large Scale Structure of the Universe (LSS) is governed by nonlinear effects of different nature: the evolution of the dark matter (DM) field, redshift space distortions (RSD), and the bias of the field for the considered tracers (galaxies, halos...) with respect to the DM one. All these effects limit the application of analytical techniques to rather large scales, thus excluding large part of the data from actual analyses. It is therefore remarkable that fully nonlinear statements can be made, in the form of ``consistency relations'' (CR) \cite{Peloso:2013zw, Kehagias:2013yd}. These are statements about the effect of perturbations at large scales on small scales ones, expressed in terms of relations between correlation functions of different order. 

The content of the CR's is essentially kinematic: it describes the effect of a long wavelength displacement field on short distance fluctuations. In linear perturbation theory, assuming adiabatic and gaussian initial conditions, the displacement field is directly related to a long wavelength matter density fluctuation, with a coefficient growing as $1/q$ as $q\to 0$, where $q$ is the long wavelength wavenumber. Since a uniform displacement cannot have any consequence on equal-times correlation functions \cite{Scoccimarro:1995if}, CR's exhibit a charachteristic $1/q$ growth only when considering unequal time correlators. On the other hand, for equal-time correlators the behavior of correlation functions as prescribed by CR's is in general `obscured' by other terms, whose form is not dictated by the CR's and are formally of the same order in $q$. Therefore the practical use of CR's appears to be of a limited extent, as unequal-time correlation functions are not measurable (notice that ``unequal-times" does not mean ``unequal redshifts", as the latter involve density fields in regions inside the past lightcone, which is not the case for the former).

In this paper we show that BAO's provide a way to disentangle the CR-protected terms from the unprotected ones also for equal-times correlators. Indeed, the BAO oscillation feature is amplified in the CR protected terms, which can then be separated from the unprotected ones, which are mostly smooth. This opens the way to the possibility of using the bispectrum CR's on real data and to extract the cosmological information imprinted on the CR coefficients. The main information we can extract is the large scale cross correlator between the velocity field and the density field for a given tracer.  Therefore, by the CR's we can extract large scale bias for that tracer in a completely model independent, and computational scheme independent, way. In this paper we present this approach by applying it on large volume N-body simulations, validating the use of CR's on the matter and halo density fields.  

The paper is organized as follows. In Sect.~\ref{CRBAO} we derive the equal time CR's and check them both in linear perturbation theory and against simulations. In Sect.~\ref{EHB} we apply the CR's to extract the bias of halos at different redshift, while in Sect.~\ref{discussion} we compare our approach with the standard perturbative one, and give our conclusions. In Appendix~\ref{derive} we give details on the derivations of the equal time CR's considered in this paper.

\section{\label{CRBAO}Consistency relations and BAO's}
The CR's are based on two ingredients. At the dynamical level, the EP, which states that a change in the phase space comoving coordinates from ($\bx$, $\bp$)  to ($\bx'$, $\bp'$), with $\bx'=\bx+\bd(\tau)$ and $\bp'=\bp+ a m \dot\bd(\tau)$, can always be absorbed by a change in the gravitational force from $\nabla \phi(\bx, \tau)$ to
\beq
 \nabla\phi'(\bx',\tau)= \nabla\phi(\bx,\tau) -{\cal H} \dot\bd(\tau) -\ddot\bd(\tau)\,,
\eeq
where $\tau$ is conformal time, $\bd(\tau)$ is an arbitrary uniform but time-dependent displacement, dots denote derivatives wrt $\tau$, and ${\cal H}=\dot a/a$.  We stress that this is an invariance of the Vlasov equation, which describes the phase space evolution  beyond the fluid approximation commonly advocated in analytical approaches, such as Perturbation Theory (PT). Therefore, the resulting CR's hold not only at all PT orders but also beyond that, including all possible non-perturbative effects such as shell-crossing and multistreaming \cite{Pietroni:2018ebj}.

The second ingredient leading to CR's comes from relating the displacement $\bd(\tau)$ to actual long wavelength velocity modes of the Universe we live in. 
The connection  is done, in Fourier space, by considering a wavenumber dependent displacement, 
\beq
\tilde\bd(\bq,\tau)=\int^\tau d\tau'\,\bv(q,\tau') = \frac{\bv(q,\tau)}{{\cal H} f} = i\frac{\bq}{q^2}\,\delta_m(\bq,\tau)\,,
\label{ldisp}
\eeq
where $\delta_m(\bq,\tau)$ is the DM overdensity field, and we have used linear PT, assuming it holds small $q$ limit. 
The effect of a long wavelength velocity field on short scale perturbations is encoded in the {\it{squeezed limit} }of the  BS
\beqra
 &&\!\!\!\!\!\!\!\!\!\!\! B_{\alpha\beta\gamma}(q,k_+,k_-;\tau_\alpha,\tau_\beta,\tau_\gamma) \equiv \nonumber\\
 &&\qquad\qquad\qquad\langle \delta_\alpha(\bq;\tau_\alpha) \delta_\beta(-\bk_+;\tau_\beta) \delta_\gamma(\bk_-;\tau_\gamma)  \rangle^\prime\,,
 \label{bisp}
 \eeqra
  where $\bk_\pm=\bk\pm\frac{\bq}{2}$, $q=|\bq|$, $k_\pm=|\bk_\pm|$,  and the prime indicates that the expectation value has been divided by a $(2\pi)^3 \delta_D(0)$ factor. $\delta_{\alpha, \beta, \gamma}$ indicate the density contrasts for different tracers (e.g. DM, baryons, a given galaxy type, ...), evaluated at times $\tau_{\alpha,\beta,\gamma}$, respectively.

 Assuming gaussian and adiabatic initial conditions in the growing mode,  the BS in the squeezed limit, $q\ll k$, receives a contribution from the long wavelength displacement modes \re{ldisp}, given by \cite{Peloso:2013spa, Baldauf:2015xfa},
 \begin{widetext} 
 \beq
 B_{\alpha\beta\gamma}(q,k_+,k_-;\tau_\alpha,\tau_\beta,\tau_\gamma) \simeq \frac{\bk\cdot\bq }{q^2}    P_{\alpha m}(q;\tau_\alpha,\tau_\alpha) \left[ \frac{D(\tau_\beta)}{D(\tau_\alpha) }P_{\beta \gamma}(k_-;\tau_\beta,\tau_\gamma)-\frac{D(\tau_\gamma)}{D(\tau_\alpha) }P_{\beta \gamma}(k_+;\tau_\beta,\tau_\gamma)
\right]+ O\left(\left(\frac{q}{k}\right)^0\right)\,,
\label{CR_diff_t}
 \eeq
 \end{widetext} 
 where the power spectra are defined as
 \beq
 P_{\alpha\beta}(k;\tau_\alpha,\tau_\beta)\equiv\langle \delta_\alpha(\bk;\tau_\alpha)\delta_\beta(-\bk;\tau_\beta) \rangle^\prime\,,
 \eeq
 $D(\tau)$ is the linear matter growth factor and we have assumed the linear behavior at the {\it soft} scale $q$,  $P_{\alpha m}(q;\tau_\alpha,\tau_\beta)=P_{\alpha m}(q;\tau_\alpha,\tau_\alpha) D(\tau_\beta)/D(\tau_\alpha)$. On the other hand,
as we have already  emphasised, the dynamics at the {\it hard} scale, $k$ is completely nonlinear.  The key point is that the structure of the first term at the RHS is protected against any kind of, perturbative and nonperturbative, nonlinear effect. By contrast, the form of the remaining terms, indicated as $O((q/k)^0)$, is not protected and will be modified in a less and less controllable way at increasing $k$ vaules and decreasing redshifts.

Considering equal times and equal species ($\alpha=\beta=\gamma$), and defining the bias parameter as
\beq
b_\alpha(q;\tau_\alpha)\equiv\frac{P_{\alpha\alpha}(q;\tau_\alpha,\tau_\alpha)}{P_{\alpha m}(q;\tau_\alpha,\tau_\alpha)}.
\label{bias_def}
\eeq
we can rewrite the squeezed limit BS in terms of the logarithmic derivative of the PS, as
\beqra
&& \lim_{q/k\to 0} \frac{B_{\alpha\alpha\alpha}(q,k_+,k_-)}{P_{\alpha\alpha}(q)P_{\alpha\alpha}(k)} =\nonumber\\
&&  - \frac{ \mu^2}{b_\alpha(q)}  \frac{ d\log P_{\alpha\alpha}(k)}{d \log k}
+ O\left(\left(\frac{q}{k}\right)^0\right),
\label{fCR}
\eeqra
where $\mu\equiv \hat\bk\cdot\hat\bq$, and we have omitted the time dependence. 
Notice that, as expected, and verified recently in N-body siumulations \cite{Esposito:2019jkb}, the equal-time CR's contains no $1/q$ pole.
Finally, we stress that we define bias as the physical quantity  in Eq.~\re{bias_def}, and not as a parameter in a given bias model, as for instance $b_1$ in the usual bias expansion $\delta_h=b_1\,\delta_m+b_2\,\delta_m^2+ b_s s^2\cdots$, (where $s^2$ is the ``tidal bias'' parameter, see e.g., \cite{Baldauf:2012hs}), see the discussion in Sect.~\ref{discussion}.
\subsection{Check in Perturbation Theory}

\begin{figure}
\includegraphics[width=0.4\textwidth]{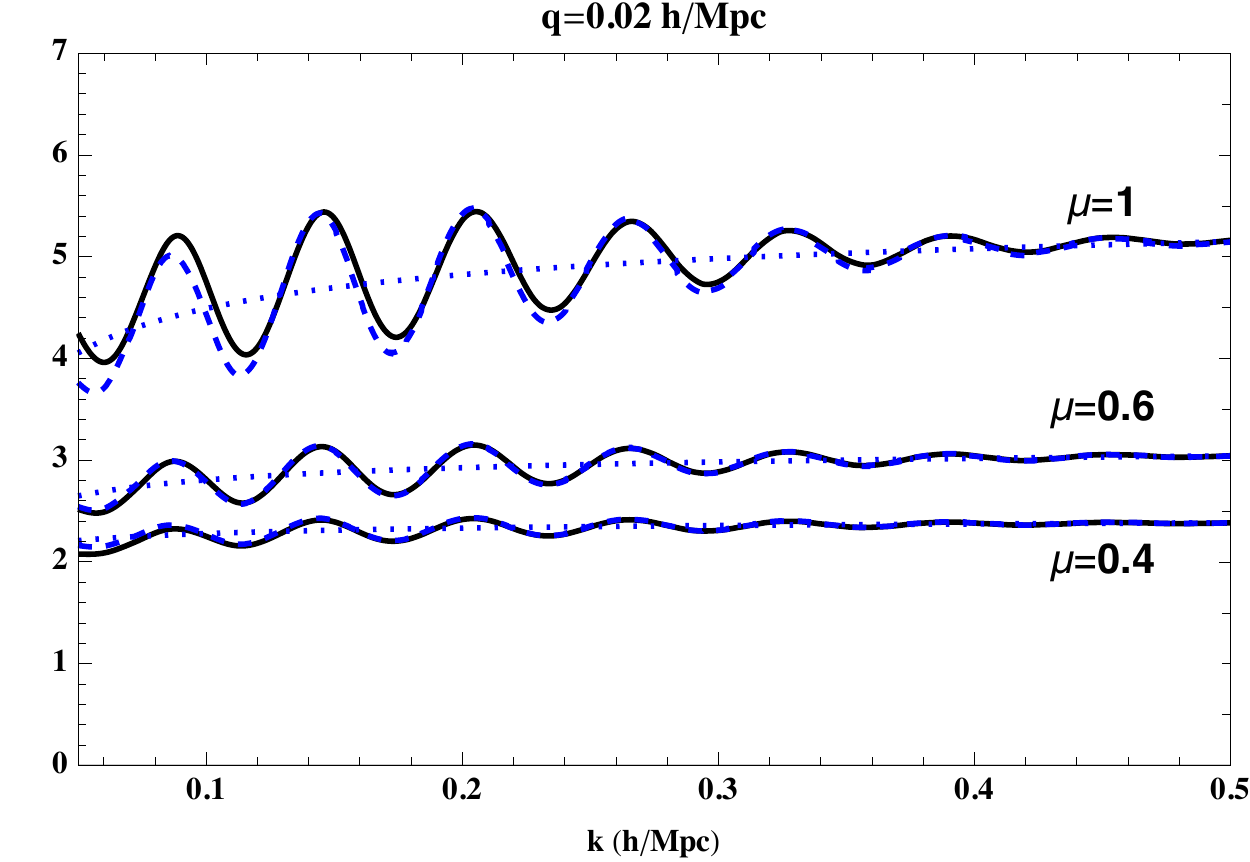}
 \caption{Comparison between the tree-level BS (divided by $P^0(q) P^0(k)$) (black-solid lines) and the RHS of Eq.~\re{crel1loop} (blue-dashed lines) as a function of $k$ for two values for $q$.  The blue-dotted lines are obtained by using the ``no-wiggle" linear power spectrum, $P^{\rm nw}(k)$, in place of $P(k)$.}
 \label{fig1}
\end{figure}
As a first test of the CR, we compute the matter BS in the squeezed limit at lowest order in SPT. It is given by 
\beqra
&&\!\!\!\!\!\!\!\!\!\!\!\!\!\! \!\!\!\!\!\!\!\!\! \lim_{q/k \to\, 0} \frac{B^{\rm SPT}_{mmm}\left(q,k_-,k_+\right)}{P^0_{m}(q)P^0_{m}(k)} = - \mu^2 \frac{ d\log P^0_{m}(k)}{d \log k}\nonumber\\
&&\;\;\,\;\, \;\qquad\qquad\qquad+\frac{13+8 \mu^2}{7} +
{\rm{O}}\left(\frac{q}{k}\right)\,,
\label{crel1loop}
\eeqra
where $P^0_{m}\equiv P^0_{mm}$ denotes the linear matter PS.  Notice that the second term at the RHS is scale independent, as it is proportional to  $(P^0_{m}(k_+)+P^0_{m}(k_-))/P^0_{m}(k)\to 2$. On the other hand, the first term, although subdominant, can be isolated from the rest thanks to its scale dependence, which is  induced mainly  by BAO oscillations. To see this, one writes the  PS as
\beq
P^0_m(k)=P^{nw}_m(k)(1 + A(k)\sin(k r_s)),
\label{oscillating}
\eeq
where $r_s$ is the comoving sound horizon and both $P^{nw}_m(k)$ and $A(k)$ are featurless functions. Performing the logarithmic derivative, we get  the first term at the RHS of \re{crel1loop} as
\beqra
&&\!\!\!\!\!\!\!\!\!-  \mu^2 \Big[ \left(  \cos\left(k r_s\right) +\alpha(k) \sin\left(k r_s\right)\right)\, \frac{ k r_s A(k) }{1+A(k) \sin\left(k r_s\right) }\nonumber\\
&&\qquad\qquad\qquad\qquad\qquad\qquad+ \frac{d \log P^{nw}_m(k)}{d\log k}\Big]  \,,
\label{spt_amp}
\eeqra
where we defined
\beq
\alpha(k)\equiv\frac{1}{k r_s } \frac{d \log A(k)}{d\log k}\,,
\eeq
which takes values around $10\,\%$ in the $k$ range of interest. So, the squeezed BS contains an oscillating component whose amplitude is enhanced by $ k r_s\simeq 2\pi k/(0.05 \,\mathrm{h \,Mpc^{-1}})$ with respect to  Eq.~\re{oscillating}  and whose phase is  shifted by $\sim \pi/2-\alpha(k)$, and a smooth component given by the second line of Eq~\re{spt_amp}.
The result is shown in Fig.~\ref{fig1}, for $q=0.02 \,\mathrm{h \,Mpc^{-1}}$ and  $q=0.05 \,\mathrm{h \,Mpc^{-1}}$ and for three different values for $\mu$. We see that the CR (Eq.~\re{crel1loop})  reproduces the BS as long as the squeezed limit ($q/k\ll 1$) holds. When $q/k$ is not small, extra scale dependent terms come into play. However, the amplitudes of the oscillations in the RHS and in the LHS are still related by the CR, as we will show quantitatively in the next section.

\subsection{Check in Simulations}

 Once nonlinear effects are included, the CR \re{fCR} ensures that  the first line in Eq.~\re{crel1loop} is modified just by changing the linear PS with the nonlinear one, while the second line is changed in an uncontrolled way. However, due to parity invariance, the BS is symmetric under $k_+\leftrightarrow k_-$, or, equivalently, $\mu\to-\mu$, which implies that the leading non-protected term in the squeezed limit should be proportional to $(P_m(k_+)+P_m(k_-))/P_m(k)$, that is, still featureless,  although with an unknown coefficient.

 To check this explicitly in fully nonlinear dynamics, we use 10 realizations of $N$-body simulations with $N=2048^3$ mass elements performed in periodic cubes with the side length of $4\,h^{-1}\mathrm{Gpc}$. The error bars presented in this paper are obtained from the scattering among the 10 realizations.
 
 The mass distribution evolved with the \textsc{Gadget2} code \cite{Springel:2005mi}  and the halo catalogs extracted by \textsc{Rockstar} algorithm \cite{Behroozi:2011ju}  at $z=0$  will be presented in what follows (see the ``fiducial cosmology'' in \cite{Nishimichi:2018etk} for other cosmological/numerical parameters). We measure the bispectrum using the quick FFT-based algorithm presented in \cite{Baldauf:2014qfa}  with the aliasing correction following \cite{Sefusatti:2015aex}.  We first store the data in bins of $(k,q,\mu)$ and check the results after summing up $B(q,k,\mu)/[P(q)P(k)]$ for different $q$ bins up to some $q_\mathrm{max}$ weighting by the number of triangles.

We compare the oscillating part of the squeezed BS to that of the logarithmic derivative of the PS, and check if their amplitudes are related by the $-\mu^2/b_\alpha(q)$ factor of  Eq. \re{fCR}. In order to extract the oscillating part,  we will subtract  smooth functions (``$poly$" in the figures) of the form
\beq
p(k,q_\mathrm{max},\mu^2)=\sum_{i=-2}^{n}a_i(q_\mathrm{max},\mu^2)k^i\,.
\label{poly}
\eeq
The negative powers in $k$, at each fixed $q_\mathrm{max}$, account for subleading terms in the squeezed limit, of order  up to $q_\mathrm{max}^2/k^2$, while the positive powers account for the extra scale dependence induced for instance by the $d \log P^{nw}_m/d\log k$ contribution.
The coefficients $a_i(q_\mathrm{max},\mu^2)$ are fixed such that $p(k,q_\mathrm{max},\mu^2)$ is the best fit to the chosen data (BS or logarithmic derivative); then the function is subtracted from the data to obtain only the oscillatory part, which is not captured by the fit for small enough $n$.
We truncate our fit at $n=2$, since with this value a satisfactory reduced $\chi^2$ (see below) is obtained. 
\begin{center}
 \begin{figure}[h]
 \includegraphics[width=0.45\textwidth]{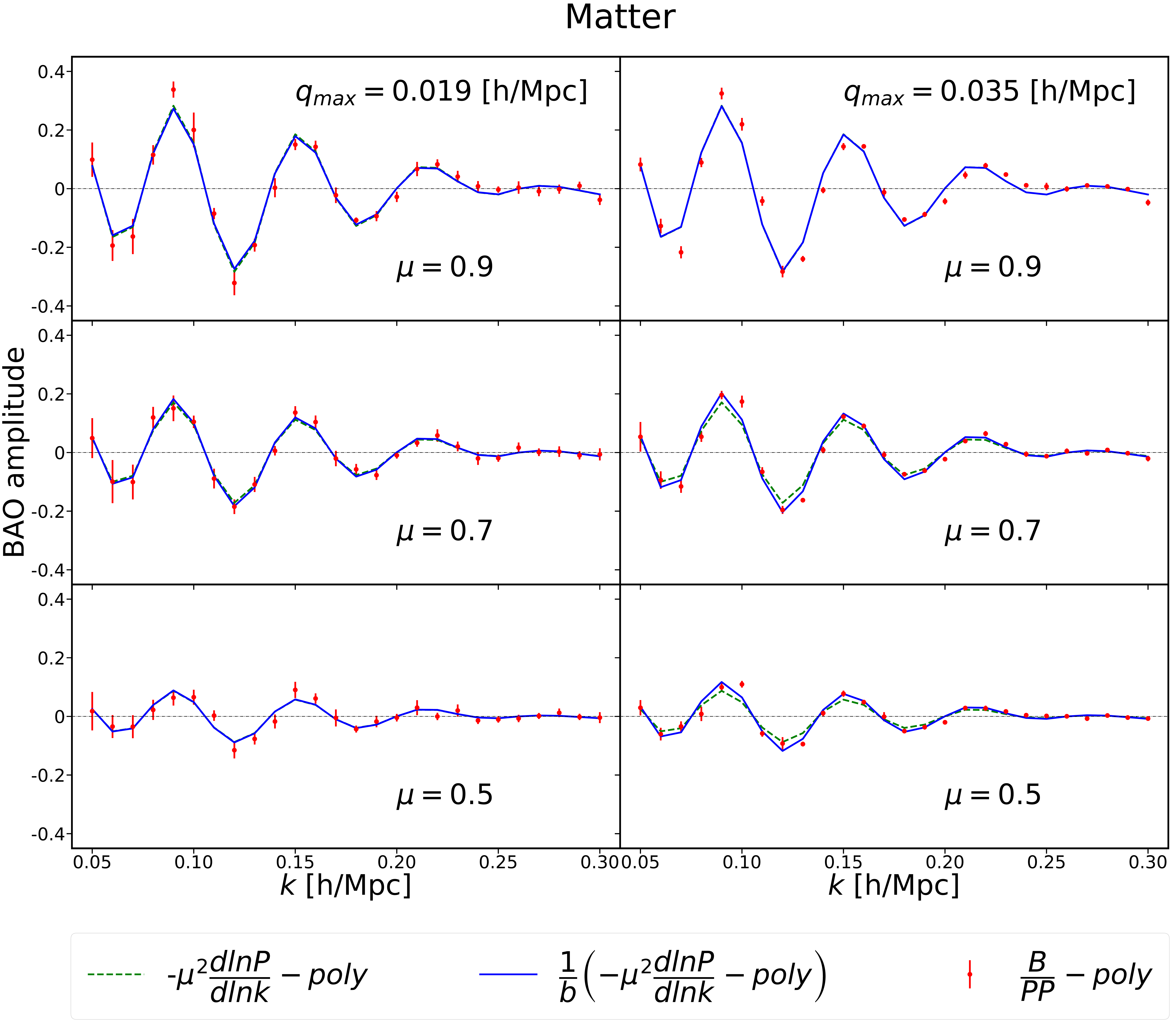}
 \caption{Comparison of the oscillating parts of the two sides of the CR, Eq. \re{fCR}, for different $\mu$ bins and different $q_{max}$,  for matter at redshift $z=0$. The red dots correspond to the BS, the blue lines are the oscillating part of the logarithmic derivative of the PS multiplied by $-\mu^2/b_m$, with $b_m$ the best fit value of tab.~\ref{table_matter}. The green-dashed lines are computed assuming the expected value, $b_m=1$.}
 \label{matter_b}
 \end{figure}
 \end{center}
To quantify the goodness of the CR we measure the bias $b_\alpha(q_{max})$ in each $\mu$ bin, by minimizing the  $\chi^2$ function
\beq
\chi^2=\sum_{i=1}^{N}\frac{\left(\frac{B}{PP}|_{i} - p_{B_i}\right) - \frac{1}{b_\alpha}\left( - \mu^2 \frac{d \ln P}{d \ln k}|_i - p_{P_i}\right)}{\sigma_i^2},
\label{chi_sq}
\eeq
where all the $q_{max}$, $k$ and $\mu$ dependencies are omitted. We denote with $p_{B_i}$ and $p_{P_i}$ the fitting curves relative to, respectively, the LHS and the RHS of the CR. $\sigma_i$ are the errors for the BS in the $i-th$ $k$-bin, since the errors on the power spectra are much smaller. We use $N=26$ linearly spaced bins from $k_{min}=0.05$ h Mpc$^{-1}$ to $k_{max}=0.30$ h Mpc$^{-1}$. 

First, we performed the test on matter, for which the expected value is $b_m(q_\mathrm{max})=1$. In Fig.~\ref{matter_b} we show the oscillating part of the BS (red points with error bars, estimated from the scatter among the 10 realizations) and of the logarithmic derivative of the PS multiplied by $-\mu^2/b_m$ (blue lines), for different  $\mu$-bins (rows) and for two different values for $q_{max}$ (columns). In
tab.~\ref{table_matter} we give the best fit values for $b_m$ in the different  $\mu$-bins and the relative $1-\sigma$ errors and reduced $\chi^2$. As expected, going to higher $q_{max}$ the quality of the fits gets worse, both because the squeezed limit is farther, and because the error bars are smaller since the number of BS configuration increases. The best fit values are  always compatible with $b_m=1$ for $q_{max}\alt 0.019 \;\mathrm{h/Mpc}$.

\begin{center}
\begin{table}[h]
\begin{tabular}{| c | c | c | c |} 
\hline
\multicolumn{4}{| c |}{$q= 0.019$ h/Mpc}\\
\hline
$\mu$ & $b_m$ & $\sigma_b$ & $\tilde{\chi}^2$\\ [0.5ex] 
\hline\hline
0.9 & 1.04 & 0.05 & 1.28\\
\hline
0.7 & 0.94 & 0.05 & 0.56\\
\hline
0.5 & 1.0 & 0.1 & 0.78\\
\hline
0.3 & 1.35 & 0.47 & 0.65\\
\hline
\end{tabular}
\begin{tabular}{| c | c | c | c |}
\hline
\multicolumn{4}{| c |}{$q = 0.035$ h/Mpc}\\
\hline
$\mu$ & $b_m$ & $\sigma_b$ & $\tilde{\chi}^2$\\ [0.5ex]
\hline\hline
0.9 & 1.00 & 0.06 & 8.68\\
\hline
0.7 & 0.84 & 0.04 & 4.22\\
\hline
0.5 & 0.74 & 0.04 & 1.91\\
\hline
0.3 & 0.8 & 0.1 & 0.84\\
\hline
\end{tabular}
\caption{Best fit values for $b_m$ (expected value, $b_m=1$) at $z=0$ for different values of the maximum allowed $q$. }
\label{table_matter}
\end{table}
\end{center} 
\section{Extracting Halo bias}
\label{EHB}
Having tested the CR on the matter field, we then used them to measure the bias of a given halo population, $b_h(q_\mathrm{max})$, via the $-\mu^2/b_h(q_\mathrm{max})$ modulation of the BAO oscillations, and  compare the results with those obtained via the definition in Eq.~\re{bias_def}. We considered halos of masses  $M>M_{min}=10^{13}M_{\odot}$ at $z=0$. The expected bias, as measured from Eq.~\re{bias_def}, is  $b_{h}=1.46\pm0.03$.
\begin{center}
 \begin{figure}
  \includegraphics[width=0.45\textwidth]{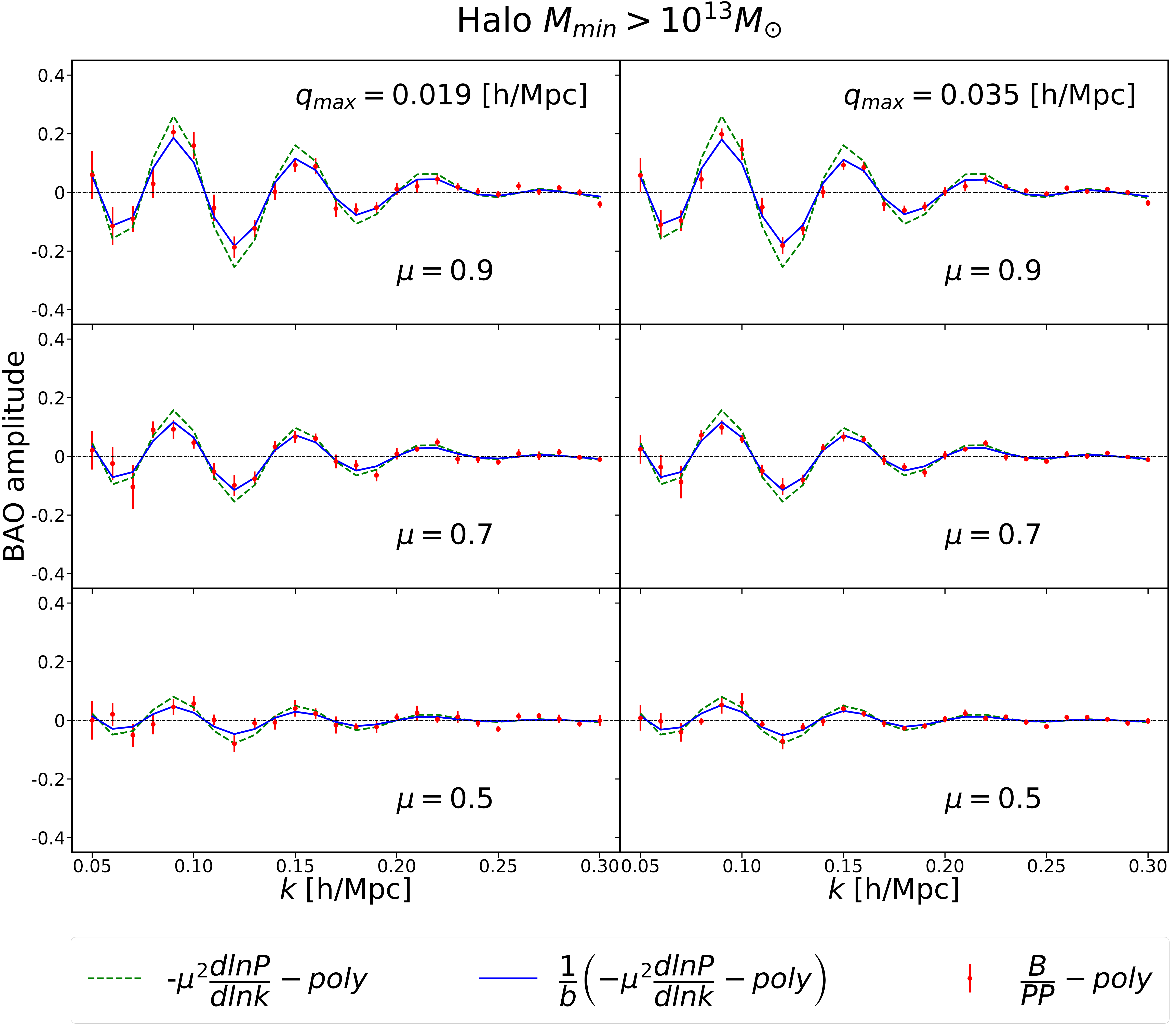}
 \caption{Same as Fig.~\ref{matter_b} for halos of mass $M>M_{min}=10^{13}M_{\odot}$ at $z=0$.}
 \label{prova_halo}
 \end{figure}
\end{center}

\begin{center}
 \begin{figure}
  \includegraphics[width=0.45\textwidth]{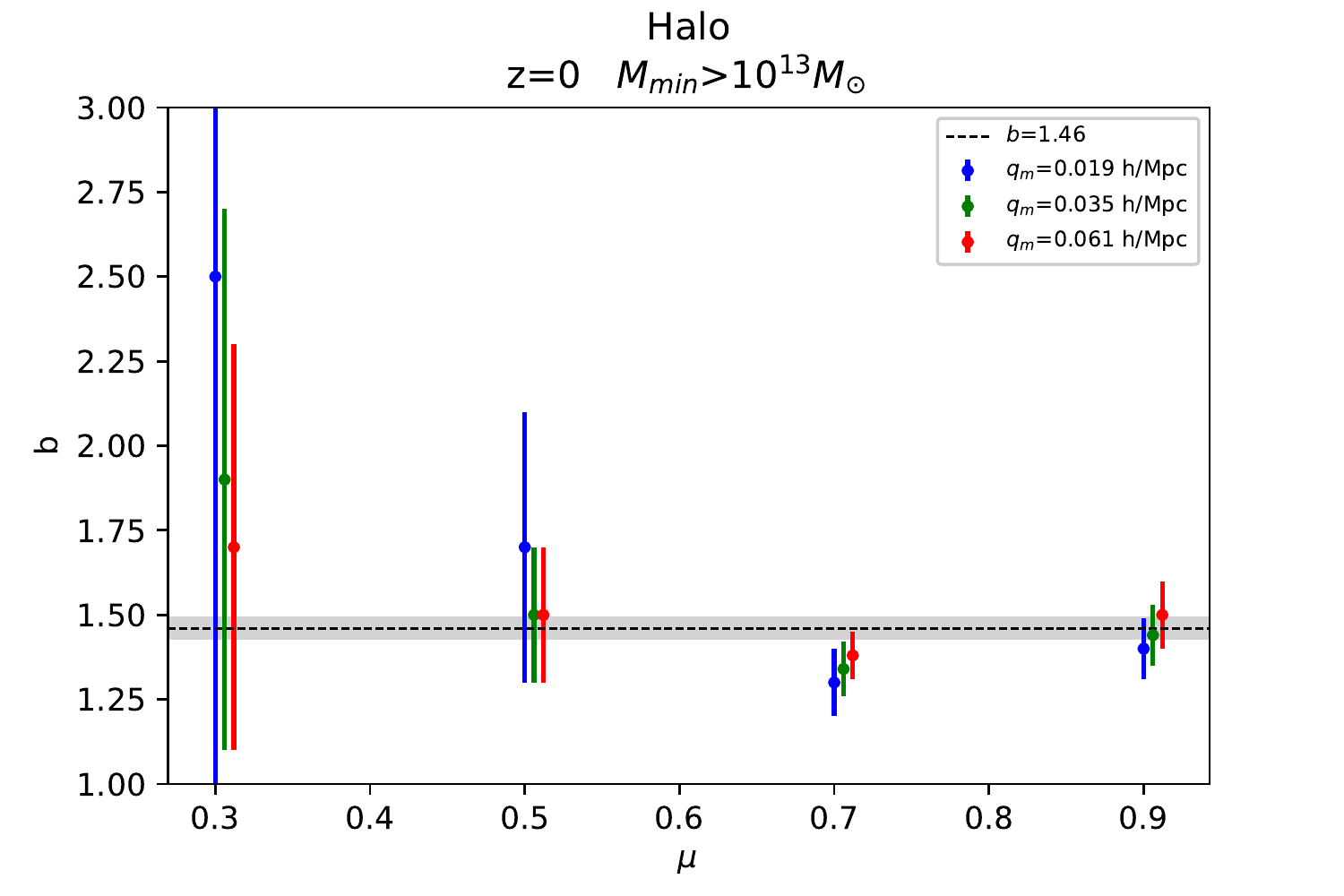}
 \caption{Values of the bias extracted for halos of mass $M>M_{min}=10^{13}M_{\odot}$ at $z=0$ for different $\mu$-bins and for different values for $q_{\rm max}$. The shaded area represents the values extracted from the simulations using Eq.~\re{bias_def} at $1$-$\sigma$.} 
 \label{halo_bias_z0}
 \end{figure}
\end{center}

\begin{center}
\begin{table}[h]
\begin{tabular}{| c | c | c | c | } 
\hline
\multicolumn{4}{| c |}{$q= 0.019$ h/Mpc}\\
\hline
$\mu$ & $b_h$ & $\sigma_b$ & $\tilde{\chi}^2$\\ [0.5ex] 
\hline
0.9 & 1.40	 & 0.09 & 0.8\\
\hline
0.7 & 1.3 & 0.1 & 0.6\\
\hline
0.5 & 1.7& 0.4 & 1.1\\
\hline
0.3 & 2.5 & 1.9 & 0.6\\
\hline
\end{tabular}
\begin{tabular}{| c | c | c | c |}
\hline
\multicolumn{4}{| c |}{$q= 0.035$ h/Mpc}\\
\hline
$\mu$ & $b_h$ & $\sigma_b$ & $\tilde{\chi}^2$\\[0.5ex]
\hline
0.9 & 1.44  & 0.09 & 1.1\\
\hline
0.7 & 1.34 & 0.2 & 0.6\\
\hline
0.5 & 1.5& 0.2 & 1.0\\
\hline
0.3 & 1.9 & 0.8 & 0.6\\
\hline
\end{tabular}
\begin{tabular}{| c | c | c | c |}
\hline
\multicolumn{4}{| c |}{$q= 0.061$ h/Mpc}\\
\hline
$\mu$ & $b_h$ & $\sigma_b$ & $\tilde{\chi}^2$\\[0.5ex]
\hline
0.9 & 1.5  & 0.1 & 1.6\\
\hline
0.7  & 1.38 & 0.07 & 0.6\\
\hline
0.5 & 1.5& 0.2 & 1.1\\
\hline
0.3 & 1.7 & 0.6 & 0.7\\
\hline
\end{tabular}
\caption{Bias values for halos with $M>M_{min}=10^{13}M_{\odot}$ at $z=0$. The expected value from Eq.~\re{bias_def} is $b_{h}=1.46\pm0.03$. }
\label{HALO_z0_13}
\end{table}
\end{center}
\begin{center}
 \begin{figure}
  \includegraphics[width=0.45\textwidth]{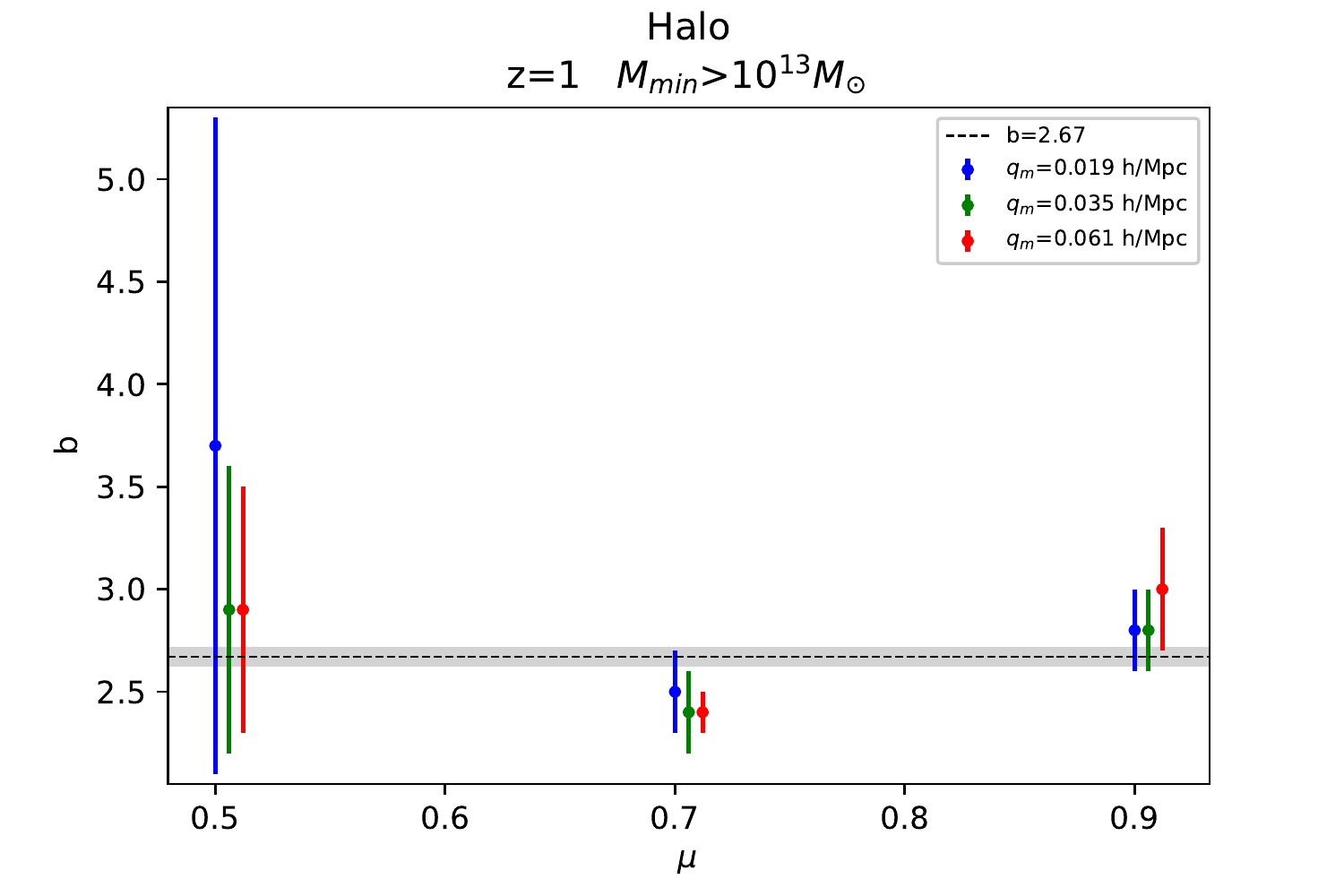}
 \caption{Same as Fig.~\ref{halo_bias_z0}  at $z=1$.}
 \label{halo_bias_z1}
 \end{figure}
\end{center}
The results are shown in Fig.~\ref{prova_halo}, where, comparing the blue lines with the green-dashed ones, we see the effect of the halo bias in reducing the amplitude of the BAO oscillations with respect to the ones present in the logarithmic derivative of the PS. As shown in Fig.~\ref{halo_bias_z0} and Tab.~\ref{HALO_z0_13}, the extracted values for $b_h$ are compatible with the expected one within the error bars, also for  $q_{max}=0.061$ h Mpc$^{-1}$, which gave bad fits in the DM case.

We have repeated the analysis for simulation data at redshift $z=1$, finding qualitatively similar results (see Fig.~\ref{halo_bias_z1}).  The only nonlinear effect limiting the effectiveness of our method  is the damping of the BAO wiggles, which is stronger at lower redshifts.

\section{discussion}
\label{discussion}
Extracting bias from a PT-based approach differs substantially from the approach discussed in this Letter. The PT prediction (at tree level as well as at higher orders) requires assuming: 1)  a cosmological model, namely, the linear matter power spectrum, its normalization $\sigma_8$ and the growth factor, and, 2) a bias model.

On the other hand, we are not assuming any model, neither for the cosmology nor for bias. Moreover, we are not relying on a computational scheme either, being it PT expansion or numerical simulations.

The crucial point is that what we call bias is not a model parameter, but a physical quantity. Thanks to the consistency relation, the cross correlator $P_{\alpha m}(q)$, and then $b_\alpha(q)$, can be measured by comparing  two directly measurable quantities, namely, the amplitude of the BAO oscillations in the logarithmic derivative of the nonlinear power spectrum and that in the bispectrum, normalized as in Eq.~\re{fCR}.
The assumptions beyond our result are just the EP, adiabatic initial conditions, and the linear continuity equation at the very large scale q, which ensure that all tracers, and in particular matter and halos, fall with the same velocity field. 
This, from a theoretical point of view, singles out our proposal as being of a qualitatively different nature, rooting its robustness in being intrinsically nonperturbative and based on very general physical assumptions.

Moreover, in a PT approach, even at tree level, three bias parameters enter, namely, the linear bias $b_1$ and the second order bias parameters, $b_2$ and  the ``tidal bias" $b_s$  (see e.g., \cite{Baldauf:2012hs}). If we try to fit the measured bispectrum with this expression, these parameters are degenerate with each other, with the normalization of the power spectrum, and possibly with other cosmological parameters.

Together with the subtleties in the treatment of the scale-dependences of the bias parameters, the end result of the fitting with the tree-level SPT template is clearly not as robust, from a theoretical point of view, as what we are proposing in this work.

\begin{center}
 \begin{figure}
  \includegraphics[width=0.35\textwidth]{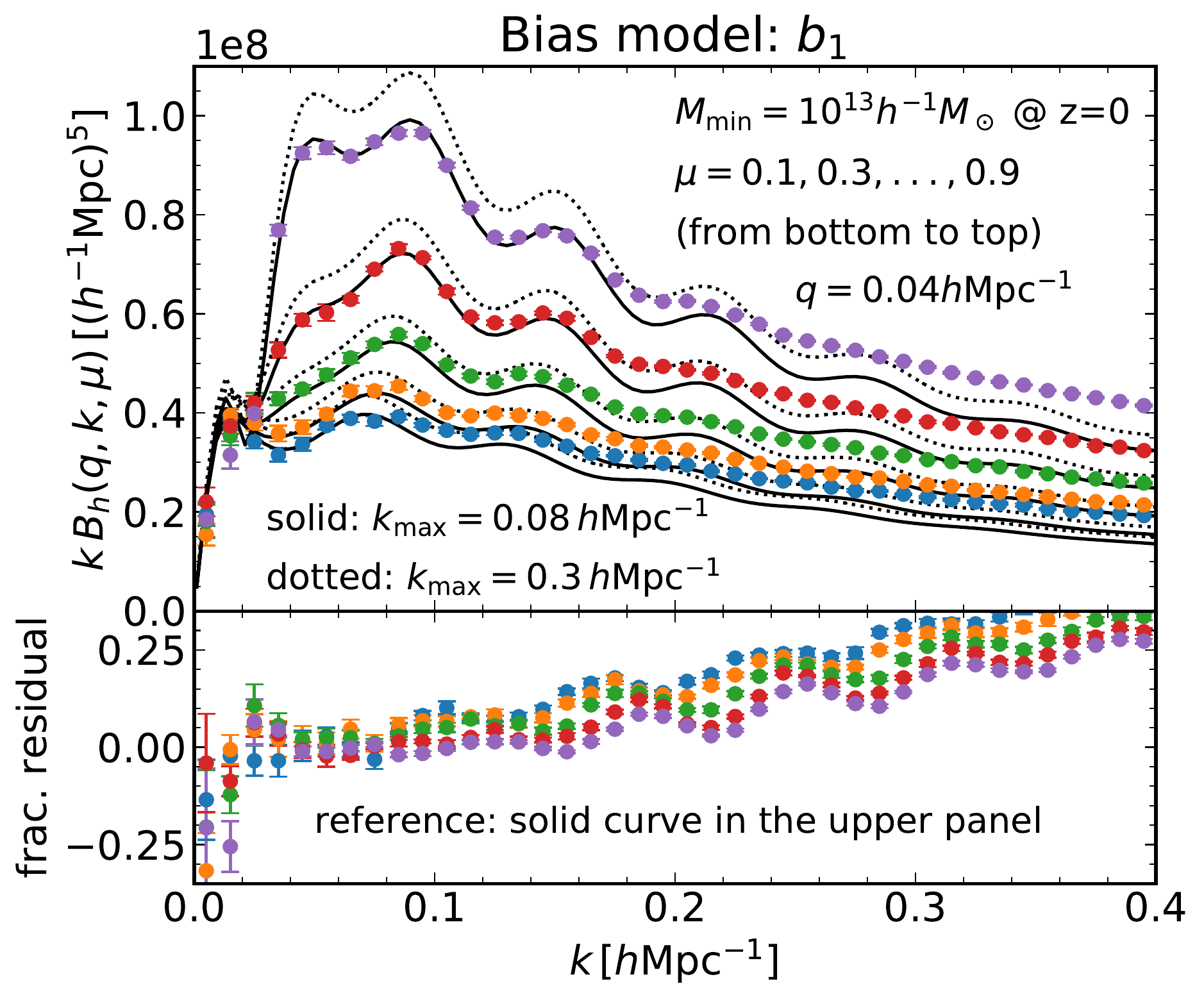}
    \includegraphics[width=0.35\textwidth]{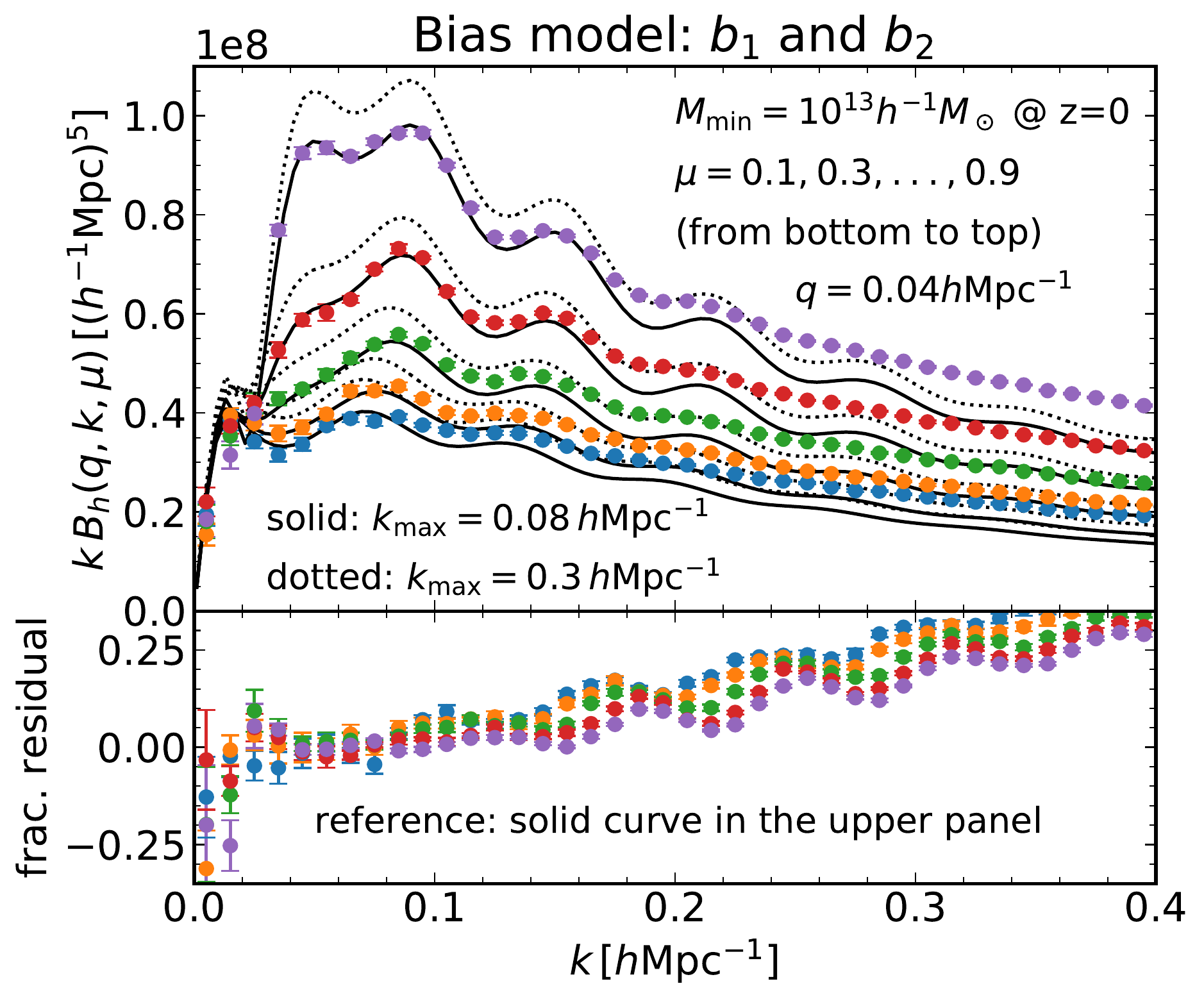}
      \includegraphics[width=0.35\textwidth]{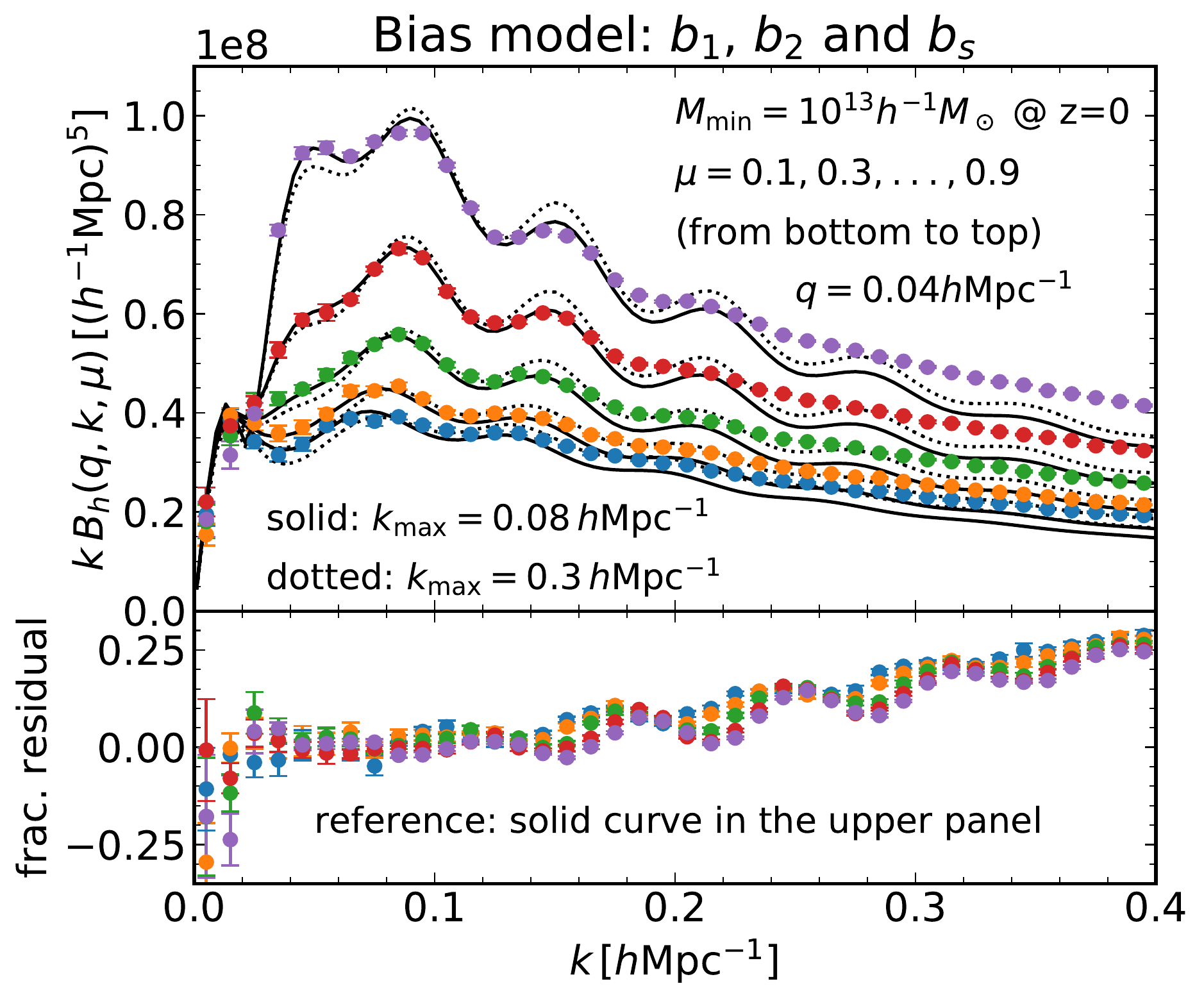}
 \caption{Fit to the full bispectrum in PT by using one, two, and three bias parameters} 
 \label{PT_fit}
 \end{figure}
\end{center}

\begin{center}
 \begin{figure}
  \includegraphics[width=0.35\textwidth]{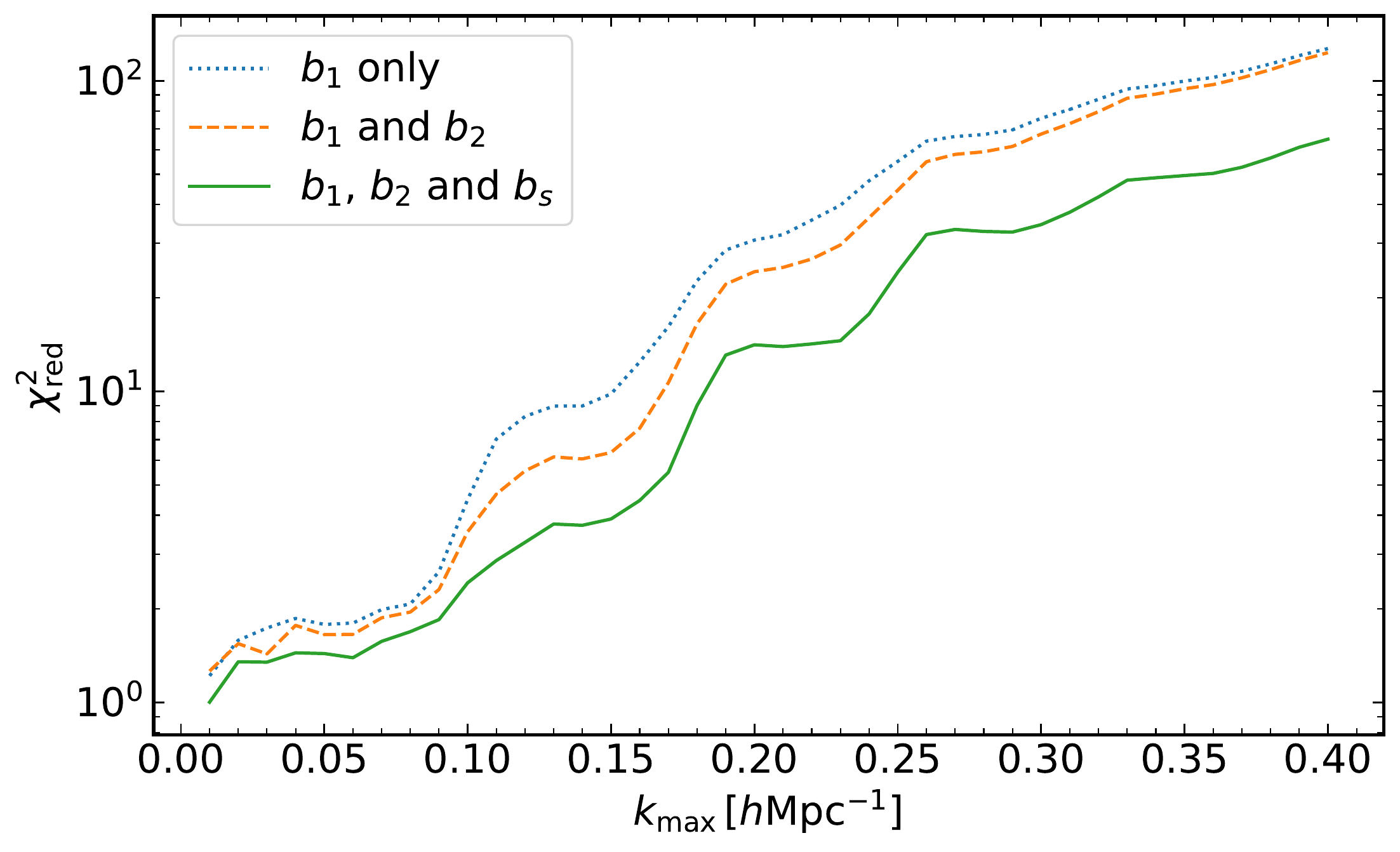}
 \caption{The reduced $\chi^2$ from the three PT fits shown in Fig.~\ref{PT_fit}, as a function of $k_{\rm max}$.} 
 \label{PTchi}
 \end{figure}
\end{center}

To see this explicitly,  we produced three plots, see Fig.~\ref{PT_fit}, in which we show the halo bispectrum at z=0 with different values of $\mu$ (in different colors) at a fixed q (=0.04 h Mpc$^{-1}$) as a function of k. The tree level calculation, but with different bias models (one for each figure, as indicated by the title) is fit to the data points, considering two different values of $k_{\rm max}$, below which the fitting was performed, $k_{\rm max}^{\rm low}=0.08$ h Mpc$^{-1}$ (solid lines), and $k_{\rm max}^{\rm high}=0.3$ h Mpc$^{-1}$ (dotted lines). We can see that, clearly, the tree-level PT does not provide a good fit over the BAO scale. We can see it quantitatively in  Fig.~\ref{PTchi}, which shows the reduced chi-squared for the three bias models as a function of $k_{\rm max}$. Indeed, it is close to unity only at very small $k_{\rm max}$.

For completeness, we summarize below the best fit bias parameters at the two values of $k_{\rm max}$ in Tab.~\ref{PTfit}.
\begin{widetext} 
\begin{center}
\begin{table}[h]
\begin{tabular}{| c | c | c | c | c| c | c |}
\hline
bias model & $b_1 (k_{\rm max}^{\rm low})$ & $b_1 (k_{\rm max}^{\rm high})$ &  $b_2 (k_{\rm max}^{\rm low})$ & $b_2 (k_{\rm max}^{\rm high})$ & $b_s (k_{\rm max}^{\rm low})$ & $b_s (k_{\rm max}^{\rm high})$\\[0.5ex]
\hline
$b_1$ & 1.347  & 1.389 & - & -&-&- \\
$b_1,\,b_2$ & 1.336  & 1.357 & 0.0355 & 0.1332&-&- \\
$b_1,\,b_2,\,b_s$ & 1.393  & 1.497 & -0.1021 & -0.3468&  -0.1371  & -0.3447 \\
\hline
\hline
\end{tabular}
\caption{Bias values for halos with $M>M_{min}=10^{13}M_{\odot}$ at $z=0$, extracted from PT fits including one, two, and three bias parameters. Results for two different  values for $k_{\rm max}$ are reported, with $k_{\rm max}^{\rm low}=0.08$ h Mpc$^{-1}$ and $k_{\rm max}^{\rm high}=0.3$ h Mpc$^{-1}$ . The expected value for $b_{h}$ defined in Eq.~\re{bias_def} is $b_{h}=1.46\pm0.03$. }
\label{PTfit}
\end{table}
\end{center}
\end{widetext} 
As we can see, the value of $b_1$, and, to a larger extent, those of $b_2$ and $b_s$, are affected by the choice of the bias model as well as the maximum wavenumber.

Moreover, we must keep in mind that these fits are performed by fixing the cosmological parameters of the PT computation to the true ones used in the simulation. In a real data analysis a scan over these parameters is needed, inducing an higher level of degeneracy.

Considering redshift space, the advantage becomes even more clear. It is known that, besides for very large scales, redshift space distortions, in particular, Fingers of God effects are particularly different to model, requiring some measure of empirical parameterization. Our approach is free from this problem, as the redshift space version of the CR assumes the Kaiser relation only where it is reliable, namely, at the large scale q, while no modelling is required at the short scale k. 

Moreover, in redshift space, the possibility of measuring, besides bias, the growth function $f=d\log D/d\log a$ opens. Indeed, following \cite{Creminelli:2013poa} , one realizes that the $\mu^2/b_h$ prefactor of Eq.~\re{fCR} now becomes
\beq
\frac{\mu+\mu_k\,\mu_q\, f}{b_h+ \mu_q^2 f} \,\mu\,, 
\eeq
where $\mu_k=\hat \bk\cdot\hat\bz$, $\mu_q=\hat \bq\cdot\hat\bz$ are the angles between the two modes and line of sight, here taken to be the $z$ axis. Therefore, by considering different orientations of the triangles, or different multipoles, or again combining different tracers, $f$ and $b_h$ can be, at least in principle, measured independently. 

We leave this analysis, as well as the quantitative assessment of the feasibility of these measurements in realistic future surveys, and the possible extension to higher order correlators for a forthcoming investigation.

\begin{acknowledgments}
We acknowledge support by the Munich Institute for Astro- and Particle Physics (MIAPP) of the DFG cluster of excellence ``Origin and Structure of the Universe''.  MP acknowledges support from the EU Horizon 2020  programme under the Marie Sklodowska-Curie grants Invisible- sPlus RISE No. 690575, Elusives ITN No. 674896. TN was supported in part by JSPS KAKENHI Grant Numbers JP17K14273 and 19H00677, and by Japan Science and Technology Agency CREST JPMHCR1414. Numerical simulations and analyses were carried out on Cray XC50 at the Center for Computational Astrophysics, National Astronomical Observatory of Japan.
\end{acknowledgments}

\iftrue
\appendix

\section{Derivation of the consistency relation}
\label{derive}
 We want to isolate the effect of nearly uniform (but time-dependent) displacements fields,
 \beq 
\bx + {\bf d}_\alpha(\bx)\,.
\label{dis}
\eeq
 on the bispectrum,
 \beq
 B_{\alpha\beta\gamma}(q,k_+,k_-) \equiv \langle \delta_\alpha(\bq) \delta_\beta(-\bk_+) \delta_\gamma(\bk_-)  \rangle^\prime\,.
 \label{bisp}
 \eeq
In order to do that, we shift back to the initial positions, defining new density fields,
 \beq
\delta_\alpha(\bx) \equiv  \tilde\delta_\alpha\left(\bx-{\bf d}_\alpha(\bx)\right)\simeq  \tilde\delta_\alpha\left(\bx\right) -{\bf d}_\alpha(\bx)\cdot \nabla  \tilde\delta_\alpha\left(\bx\right)  \,,
 \eeq
 which, in Fourier space reads  (we use the same symbol for fields in real and in Fourier space),
 \beqra
 && \delta_\alpha(\bp)\simeq  \tilde\delta_\alpha(\bp)-i\int\frac{d^3 q^\prime}{(2\pi)^3} \left(\bp-\bq^\prime\right)\cdot {\bf d}_\alpha(\bq^\prime)  \tilde\delta_\alpha(\bp-\bq^\prime).\nonumber\\
 &&
 \label{ft}
 \eeqra
 Inserting it in \re{bisp} we get two contributions containing the expectation values 
 \beq
 \langle \tilde\delta_\alpha(\bq) {\bf d}_\beta(\bq^\prime) \tilde\delta_\beta(-\bk_+ -\bq^\prime)   \tilde\delta_\gamma(\bk_-) \rangle^\prime,
 \label{ev4}
 \eeq
 and the one obtained by this via the replacements $\beta\leftrightarrow \gamma$ and $\bk_+\leftrightarrow -\bk_-$.
The CR is obtained by considering the $q\ll k$ limit, and assuming that  ${\bf d}_\beta(\bq^\prime)$ has support for $q^\prime \ll k$, and that it correlates with $ \tilde\delta_\alpha(\bq)$, that is, \re{ev4} can be written as
\beqra
&&\simeq  \langle \tilde\delta_\alpha(\bq) {\bf d}_\beta(\bq^\prime)\rangle \langle \tilde\delta_\beta(-\bk_+ -\bq^\prime)   \tilde\delta_\gamma(\bk_-) \rangle^\prime\nonumber\\
&& (2\pi)^3\delta_D(\bq+\bq^\prime)  \langle \tilde\delta_\alpha(\bq) {\bf d}_\beta(-\bq)\rangle^\prime \langle  \tilde\delta_\beta(-\bk_-)   \tilde\delta_\gamma(\bk_-) \rangle^\prime,\nonumber\\
\eeqra
and therefore the bispectrum is given, in this limit, by
\beqra
 &&\!\!\!\!\!\!\!\!\! \lim_{q/k\to 0} B_{\alpha\beta\gamma}(q,k_+,k_-) \simeq \langle \tilde\delta_\alpha(\bq) \tilde\delta_\beta(-\bk_+) \tilde\delta_\gamma(\bk_-)  \rangle^\prime +\nonumber\\
 &&\!\!\!\!\!\!\!\!\! i \bk\cdot\Big[  \langle \tilde\delta_\alpha(\bq) {\bf d}_\beta(-\bq)\rangle^\prime P_{\beta\gamma} (k_-) -  \langle \tilde\delta_\alpha(\bq) {\bf d}_\gamma(-\bq)\rangle^\prime P_{\beta\gamma} (k_+)\Big]\nonumber\\
 && +\cdots,
\eeqra
where dots indicate higher orders in the displacement fields.

Now, the EP and adiabatic initial conditions ensure that, at large scales, all species fall with the same acceleration under the effect of the potential generated by the total matter field, that is, 
${\bf d}_\beta(\bq), {\bf d}_\gamma(\bq) \to {\bf d}_m(\bq)$, as $q\to 0$. Moreover, assuming that at large scales linear theory holds, we have
\beq
{\bf d}_m(\bq) = i\frac{\bq}{q^2}\delta_m(\bq),
\eeq
and therefore the squeezed limit bispectrum can be written as
\beqra
&& \lim_{q/k\to 0} \frac{B_{\alpha\beta\gamma}(q,k_+,k_-)}{P_{\alpha\alpha}(q)P_{\beta\gamma}(k)} =\nonumber\\
&& \frac{\mu}{b_\alpha(q)} \frac{k}{q} \frac{\left(P_{\beta\gamma}(k_-)-P_{\beta\gamma}(k_+)\right)}{P_{\beta\gamma}(k)}+ O\left(\left(\frac{q}{k}\right)^0\right)=\nonumber\\
&&  - \frac{ \mu^2}{b_\alpha(q)}  \frac{ d\log P_{\beta\gamma}(k)}{d \log k}
+ O\left(\left(\frac{q}{k}\right)^0\right).
\label{fCR_app}
\eeqra
where $\mu\equiv (\bq\cdot\bk)/(q k)$, and we have defined the bias parameter as in Eq.~\ref{bias_def}.
If the fields $\delta_\beta$ and $\delta_\gamma$ were evaluated at different times, then the cancellation of the $1/q$ divergences would not be exact
  \cite{Peloso:2013zw}, see Eq.~\re{CR_diff_t}. Here there is no such divergence, still the first term in the last line of Eq.~\re{fCR} is protected by the CR, that is, the coefficient $\mu^2/b_\alpha(q)$ is not modified by any nonlinear effect, including the non-perturbative ones due to the breakdown of the fluid approximation at small scales (large $k$'s). 

It is instructive to derive  the configuration space counterpart of Eq.~\re{fCR_app}. A perfectly uniform displacement field cannot give any contribution to the equal times bispectrum. Therefore, the effect can depend only on the gradient of the large scale displacement/velocity field. More precisely, the $i-th$ component of the displacement field can affect the clustering on short scales along the $j-th$ direction only via its $\partial_j d^i(\bx)$ component, leading to a contribution to the configuration space three point function proportional to 
\begin{widetext}
\beqra
&&  \langle \delta_\alpha(-{\bf R}) \delta_\alpha\left(\frac{\bf r}{2}\right) \delta_\alpha\left(-\frac{\bf r}{2}\right)\rangle =\langle \bar\delta_\alpha(-{\bf R}+\bd(-{\bf R})) \bar\delta_\alpha\left(\frac{\bf r}{2} +\bd\left(\frac{\bf r}{2} \right) \right) \bar\delta_\alpha\left(-\frac{\bf r}{2} +\bd\left(-\frac{\bf r}{2} \right) \right)  \rangle\nonumber\\
&& \qquad\;\;\,=  r^j\frac{\partial \xi_\alpha(r)}{\partial r^i}\,  \langle \delta_\alpha(-{\bf R})\left.\frac{ \partial d^i(\br)}{\partial r^j}\right|_{\br=0}\rangle +\cdots= \frac{2}{3 {\cal H}^2}\frac{r^ir^j}{r^2}\frac{\partial \xi_\alpha(r)}{\partial \ln r}\,  \langle \delta_\alpha(-{\bf R})\left.\frac{ \partial^2 \Phi(\br)}{\partial r^i\partial r^j}\right|_{\br=0}\rangle +\cdots\,
\eeqra
\end{widetext}
where $\xi_\alpha(r)$ is the correlation function and $  \Phi(\br)$ the gravitational potential, which has been related to the displacement $\bd$ by means of linear PT.

\fi


\bibliography{/Users/massimo/Bibliografia/mybib.bib}
\end{document}